# Smaller desert dust cooling effect estimated from analysis of dust size and abundance

Jasper F. Kok[1,*], David A. Ridley[2], Qing Zhou[3], Ron L. Miller[4], Chun Zhao[5], Colette L. Heald[2,6], Daniel S. Ward[7], Samuel Albani[8], and Karsten Haustein[9]

[1]Department of Atmospheric and Oceanic Sciences, University of California, Los Angeles, CA 90095
[2]Department of Civil and Environmental Engineering, Massachusetts Institute of Technology, Cambridge, Massachusetts, United States
[3]Department of Statistics, University of California, Los Angeles, CA 90095
[4]NASA Goddard Institute for Space Studies, New York NY 10025
[5]School of Earth and Space Sciences, University of Science and Technology of China, Hefei, Anhui, China
[6]Department of Earth, Atmospheric and Planetary Sciences, Massachusetts Institute of Technology, Cambridge, MA, USA
[7]Program in Atmospheric and Oceanic Sciences, Princeton University, Princeton, NJ, USA
[8]Laboratoire des Sciences du Climat et de l'Environnement, CEA-CNRS-UVSQ, Gif-sur-Yvette, France
[9]School of Geography and the Environment, University of Oxford, Oxford OX1 3QY, UK
[*]e-mail: jfkok@ucla.edu

**Desert dust aerosols affect Earth's global energy balance through direct interactions with radiation, and through indirect interactions with clouds and ecosystems. But the magnitudes of these effects are so uncertain that it remains unclear whether atmospheric dust has a net warming or cooling effect on global climate. Consequently, it is still uncertain whether large changes in atmospheric dust loading over the past century have slowed or accelerated anthropogenic climate change, or what the effects of potential future changes in dust loading will be. Here we present an analysis of the size and abundance of dust aerosols to constrain the direct radiative effect of dust. Using observational data on dust abundance, in situ measurements of dust optical properties and size distribution, and climate and atmospheric chemical transport model simulations of dust lifetime, we find that the dust found in the atmosphere is substantially coarser than represented in current global climate models. Since coarse dust warms climate, the global dust direct radiative effect is likely to be less cooling than the ~-0.4 W/m$^2$ estimated by models in a current global aerosol model ensemble. Instead, we constrain the dust direct radiative effect to a range between -0.48 and +0.20 W/m$^2$, which includes the possibility that dust causes a net warming of the planet.**

The direct radiative effect (DRE) of desert dust aerosols on global climate depends sensitively on both the size distribution and atmospheric abundance of dust[1-3]. However, current global model estimates of the atmospheric loading of dust with geometric diameter $D \leq 10$ μm (PM$_{10}$) vary widely from ~6 to 30 Tg[4-7]. Similarly, the size distribution of atmospheric dust varies substantially across models, with the fraction of dust in the clay size range ($D \leq 2$ μm) varying by over a factor of three[8]. This uncertainty in dust size and abundance is partially driven by a critical limitation of global models: the need to prescribe poorly known attributes of dust particles. In particular, the assumed dust optical properties and size distribution at emission greatly affect the resultant size-resolved dust loading[1, 6]. Each model parameterizes these properties differently, and in a manner not always consistent with experimental results[8-10]. This divergence in assumed dust properties contributes to a wide range of estimates of the size-resolved global dust loading[6, 8]. Because fine dust cools global climate whereas coarse dust ($D \geq 5$ μm) likely warms it[3], this uncertainty in size-resolved dust loading contributes to a wide spread in model estimates of the dust DRE[1, 3, 9, 11-14].

Since the use of global models alone is thus unlikely to substantially narrow the uncertainty on dust climate effects[15], we develop an alternative approach to determine the size-resolved global dust loading, which we subsequently use to constrain the dust DRE. We use an analytical framework that leverages observational and experimental constraints on dust properties, and uses global models only where such constraints are not available. Specifically, we link dust loading to the dust aerosol optical depth (DAOD), which we constrain by combining extensive ground-based and satellite observations with global model simulations[16] (Fig. 1a). Since the globally-averaged DAOD quantifies the total extinction of solar radiation by dust in the atmosphere, we can use it to determine the dust loading if we also constrain the size distribution of atmospheric dust, and the efficiency $Q_{ext}$ with which dust of a given size extinguishes solar radiation (see Materials and Methods).

**Constraints on atmospheric dust properties and abundance**

We constrain the globally-averaged dust extinction efficiency $Q_{ext}$ (Fig. 1b) by combining experimental constraints on dust optical properties and shape with a dust single-scattering database[17]. We find that the common simplification to treat dust as spherical particles[1-3] results in an underestimation of $Q_{ext}$ by ~20–60% for dust with $D \geq 1$ μm (Fig. 1b). This underestimation is largely caused by the greater surface-to-volume ratio of irregularly-shaped dust, relative to that of an equal-volume sphere[18].

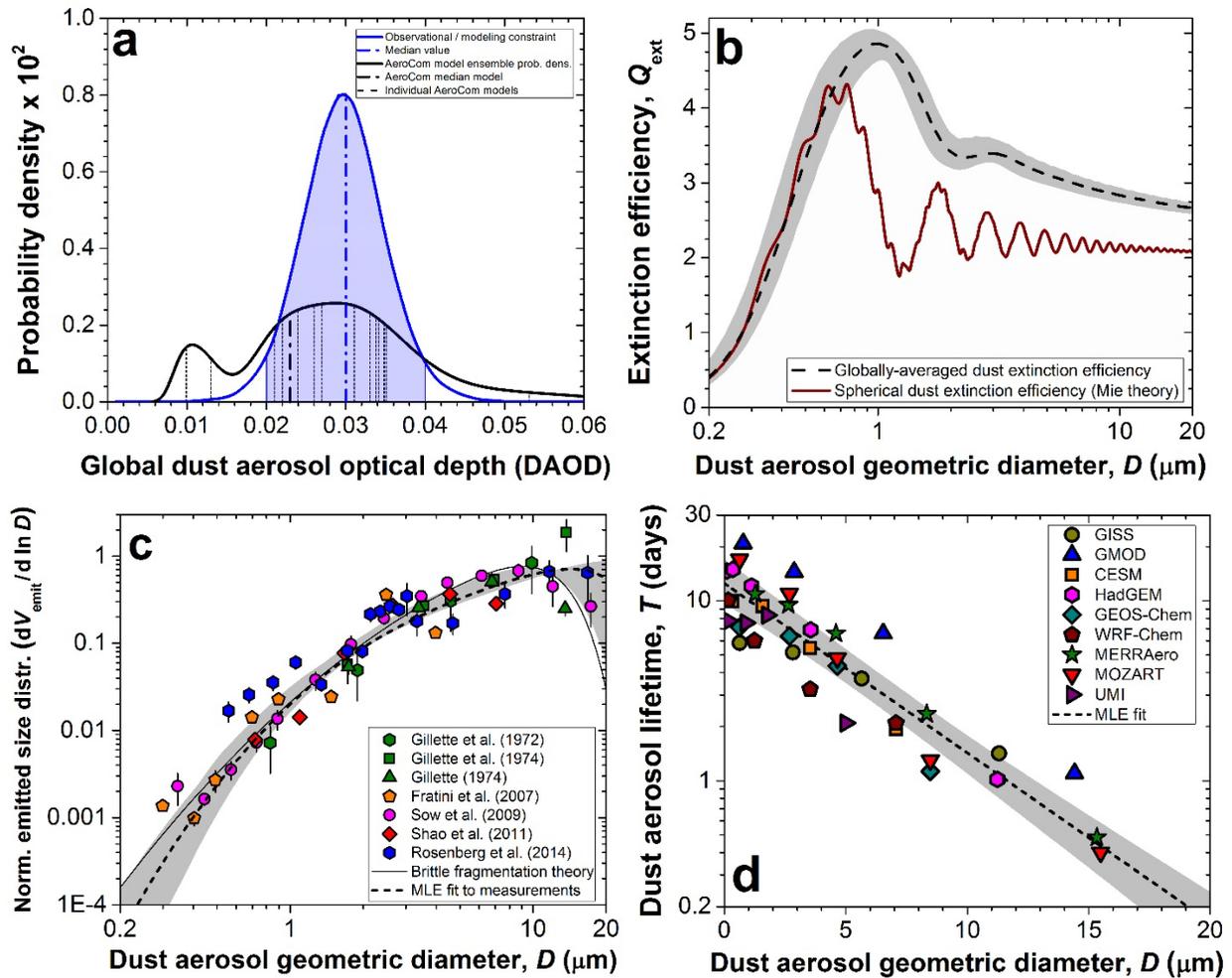

**Figure 1. New constraints on dust properties and prevalence.** (**a**) Joint observational and modeling constraint on global DAOD[20] (shading denotes 95% confidence interval (CI)), which is more precise than the AeroCom model ensemble[21]. (**b**) Joint experimental and modeling constraint on the globally-averaged dust extinction efficiency $Q_{ext}$, showing that "spherical" dust substantially underestimates $Q_{ext}$. For **b-d**, dashed lines and shading represent the maximum likelihood estimated (MLE) values and CI (see Materials and Methods). (**c**) Experimental constraint on the globally-averaged emitted dust size distribution (normalized to unity when summed over all sizes), obtained by combining five data sets in a statistical model. (**d**) Modeling constraint on the globally-averaged size-resolved dust lifetime, showing that lifetime decreases roughly exponentially with increasing dust size.

We obtain the size distribution of atmospheric dust from experimental constraints on the size distribution of emitted dust (Fig. 1c) and global modeling constraints on the atmospheric lifetime of emitted dust (Fig. 1d) (see Materials and Methods). We constrain the globally-averaged emitted dust size distribution using five data sets from a variety of dust source regions (Fig. 1c). We use a statistical model that accounts for systematic errors inherent in each study's measurement methodology, which allows us to constrain the emitted dust size distribution more strongly than otherwise possible (see Supplement for details). We find that clay-sized aerosols account for only 4.3% (95% confidence interval: 3.5–5.7%) of the emitted mass with $D \leq 20$ μm ($PM_{20}$), which is substantially less than the 5–35% assumed in global models[8]. This finding is similar to a recent result[8] based on brittle fragmentation theory (black line in Fig. 1c), which is reinforced here by the inclusion of three additional data sets. We constrain the globally-averaged size-resolved dust lifetime (Fig. 1d) using simulation results from nine global models, which we again combine using a statistical model (see Supplement). We find that the lifetime of submicron dust is 11 (9 – 15) days, and that it decreases roughly exponentially with increasing $D$. This occurs primarily because of the increase of gravitational deposition with particle diameter[3, 19]. Despite their small emitted fraction, the long lifetime of clay-sized dust causes those particles to account for 15 (12–21)% of the atmospheric mass load, and their large surface-to-volume ratio and extinction efficiency (Fig. 1b) causes them to account for about half [46 (41–56)%] of the global DAOD (Fig. S1).

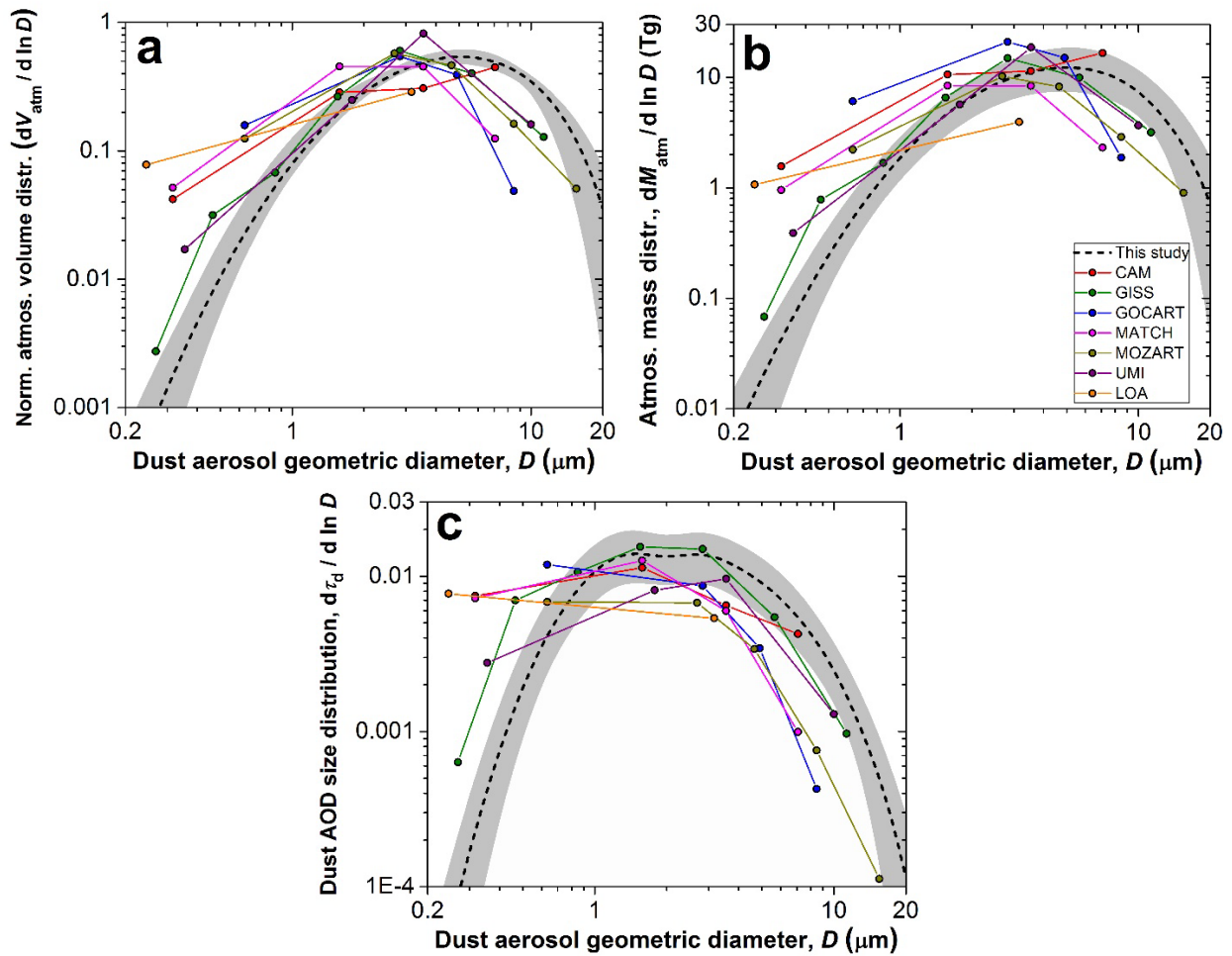

**Figure 2. Size-resolved global loading of desert dust aerosols.** (**a**) The globally-averaged normalized volume distribution (shading represents 95% CI) peaks at a coarser size than in current global models in the AeroCom ensemble[21] (colored lines). Constraints on the (**b**) size-resolved atmospheric dust mass and (**c**) the dust AOD size distribution indicate that current global models contain too much fine dust and not enough coarse dust. In contrast to the volume distribution in panel (**a**), the mass distribution is not normalized, such that its integral over size equals the global dust load.

**The size-resolved global loading of desert dust**

We obtain the normalized globally-averaged dust size distribution (Fig. 2a) by combining our constraints on the emitted dust size distribution and lifetime (see Materials and Methods). We find that dust in current global models is too fine (Fig. 2b), which is consistent with recent observations[1, 19] and was previously suggested using brittle fragmentation theory[8].

We combine the constraints on the atmospheric size distribution (Fig. 2a) with those on the DAOD (Fig. 1a) and the extinction efficiency (Fig. 1b) to obtain the global $PM_{10}$ dust emission rate $F_{emit}$ and loading $L_{atm}$ (see Materials and Methods). We find that $F_{emit} = 1.7\ (1.0 - 2.7) \cdot 10^3$ Tg/year and $L_{atm} = 20\ (13 - 29)$ Tg (Fig. 3). The global emission rate and loading of $PM_{20}$ dust are $3.0\ (1.7-4.9) \cdot 10^3$ Tg/year and $23\ (14 - 33)$ Tg, respectively (Fig. S1). Since results from the AeroCom ensemble indicate that the atmospheric loading of non-dust aerosols is around 10 Tg[5], we conclude that desert dust likely dominates global aerosol by mass. Most of the AeroCom models, as well as the median model, simulate a dust emission rate and loading below our central estimates (Fig. 3)[6], predominantly because of an underestimation of coarse dust ($D > 5$ μm; Figs. 2b and S2).

Because global models need to assume specific values for dust attributes, their results can be biased if the assigned values are not consistent with experimental results. In particular, inconsistent values for dust optical properties and the emitted particle size distribution generate biases in the size-resolved atmospheric dust loading[12,14,15], and thus in the simulated dust effects on climate[1, 3, 8]. Current models assume an emitted dust size distribution that is much finer than measurements indicate (Fig. S2), which results in a substantial bias toward fine dust in the atmosphere (Fig. 2). Since fine dust mostly scatters, whereas coarse dust also absorbs solar radiation, this fine-size bias likely contributes to the underestimation of aerosol absorption in models[22].

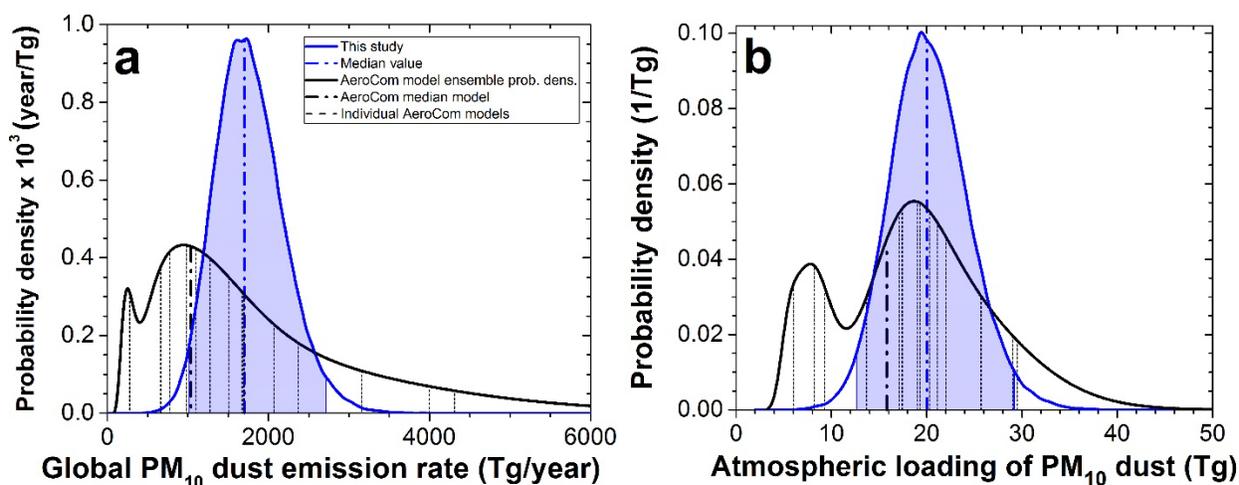

**Figure 3. Global emission rate and atmospheric loading of desert dust aerosols.** Probability densities of (**a**) the global dust emission rate and (**b**) the atmospheric dust loading (blue lines with shaded CI) indicate that some global models in the AeroCom ensemble[21] underestimate dust emission and loading.

A second bias in models results from the assumption that dust is spherical[5,15,19,20,26]. This is problematic because simplifying the highly aspherical dust particles[23] leads to a substantial underestimation of the extinction efficiency (Fig. 1b). For the atmospheric dust size distribution obtained here (Fig. 2a), the assumption of spherical dust results in an underestimation of the extinction produced by a unit mass of dust loading of 29 (24–34)%, which is consistent with recent results from deposited dust in ice cores[24]. This substantial bias is masked by excessive fine dust in models, which increases the extinction produced by a unit mass of dust (see Figs. 1d and S1). Global models furthermore slightly underestimate the global DAOD[16] (Fig. 1a). The net result of these three biases is a slight underestimation of global dust loading (Fig. 3).

**Constraints on the dust direct radiative effect**

A crucial advantage of our analytical framework is that it is subject to fewer of these biases, because it integrates observational and experimental constraints. Despite important limitations of our approach (see Materials and Methods), we consider our constraints on the size-resolved global dust emission rate and loading (Figs. 2, 3) to be more accurate and robust than constraints derived from model ensembles[4-7]. As such, our constraints on the size-resolved dust loading can better inform dust effects on climate through interactions with ecosystems[25, 26], clouds[27, 28], and radiation. The dust DRE[2, 3] is particularly sensitive to the atmospheric dust size distribution, as fine dust cools global climate by scattering solar radiation, whereas coarse dust ($D \geq 5$ μm) likely warms by absorbing both solar and thermal radiation[3] (Fig. S3). Consequently, our finding that atmospheric dust is coarser than represented in the current ensemble of global models[6] implies that dust DRE is more positive than the -0.30 to -0.60 W/m$^2$ estimated by AeroCom models[3, 9, 29, 30].

We determine the DRE of PM$_{20}$ dust by combining results on the size-resolved extinction of SW radiation (Fig. 2c) with an ensemble of model simulations of the efficiency with which a unit of extinction is converted to DRE (Fig. S3; Materials and Methods). Using the size-resolved dust loading obtained by AeroCom models yields a DRE at top-of-atmosphere (TOA) of -0.46 (-0.78 to -0.03) W/m$^2$, which is consistent with estimates by individual AeroCom models[3, 9, 29, 30] (Fig. 4). In contrast, using our constraints on the size-resolved dust loading yields a DRE of -0.20 (-0.48 to +0.20) W/m$^2$ (Fig. 4), which is consistent with recent work[13, 14] that used an emitted size distribution similar to our experimental constraints (Fig. 1c). This represents a reduction of the most likely DRE by approximately a factor of two, and a 25% chance that the global DRE is actually positive.

Three different factors contribute to our result that the dust DRE is substantially more positive (warming) than accounted for by current AeroCom models[6]. First, correcting the fine-size bias in models reduces SW cooling by ~0.15 W/m$^2$, both because fine dust predominantly scatters whereas coarse dust also absorbs, and because the short lifetime of coarse dust concentrates these particles over bright deserts, which reduces the cooling effect of scattering and enhances the warming effect of SW absorption. Second, the increase in coarse dust increases the warming arising from LW interactions by ~0.10 W/m$^2$ (Fig. 4). Finally, very coarse dust ($D >$ 10 μm) produces a positive DRE of +0.03 (+0.01 to +0.06) W/m$^2$, which is neglected by about half the AeroCom models[6].

Although our results indicate that the global dust DRE is substantially more positive than represented in current models (Fig. 4), the effects of the fine-size bias in current models are region-specific. This spatial variability in the dust DRE is primarily driven by regional differences in surface albedo and prevalence of clouds, and by the size-dependent dust lifetimes (Fig. 1d). Close to source regions, the coarse particles missing from current models produce additional warming (Fig. S4), especially over highly reflective arid regions. Further from source regions, much of this missing coarse dust has been deposited (Figure 1d and Refs. [19, 32]). However, the excess of fine dust in current models (Fig. 2b) causes an overestimation of dust cooling far from source regions (Fig. S4), particularly over low reflectivity regions, such as oceans and forests. Our results thus imply a more positive dust DRE, both close to and far from source regions.

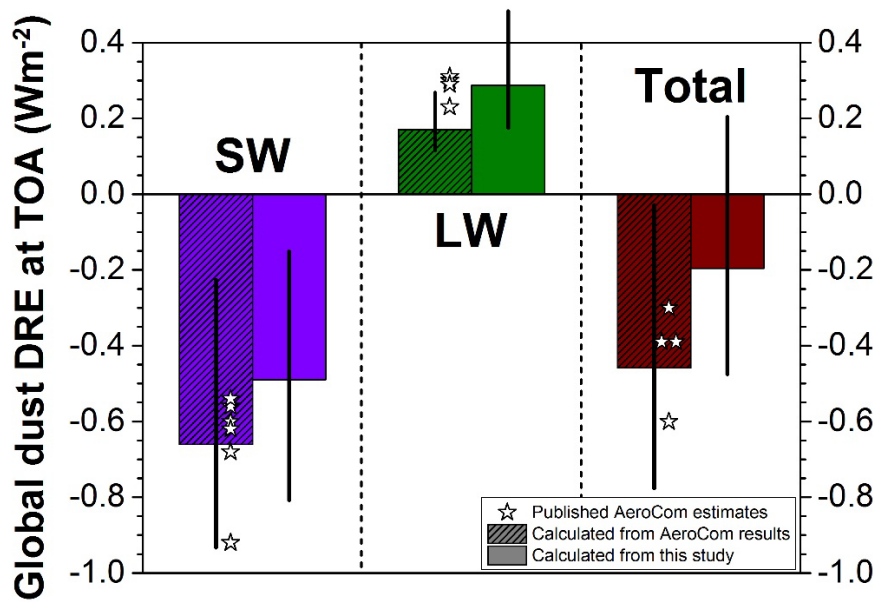

**Figure 4. Constraints on the global direct radiative effect (DRE) of PM$_{20}$ dust**. The fine-size bias in current models causes an overestimation of SW cooling and underestimation of LW warming (hatched bars). We correct these biases using our constraints on the global size-resolved dust load (Fig. 2b) and extinction efficiency (Fig. 1b), resulting in a more positive (warming) DRE at the top-of-atmosphere. Error bars denote 95% CI[9, 29-31].

Our results suggest that dust cools the climate system substantially less than represented in current models, and raise the possibility that dust is actually net warming the planet. This has important implications for the role of changes in dust loading in past and future climate changes. Past increases in dust loading[11, 33, 34] have likely slowed anthropogenic greenhouse warming less than current models suggest[11, 34], and might even have accelerated it. This is consistent with recent insights that aerosol radiative forcing might be less cooling than previously thought[15]. Similarly, anthropogenic dust emissions, which are estimated to account for about a quarter of total dust emissions[35], might enhance, rather than oppose[7], global warming. Our results further suggest that possible future increases in dust loading might dampen global climate change less than current models estimate[36], and might even enhance it.

**Additional information**
Correspondence and requests for materials should be addressed to J.F.K. (jfkok@ucla.edu).

**Acknowledgements**
We thank Virginie Bouchard, Bingqi Yi, Kuo-Nan Liou, Ping Yang, Aradhna Tripati, David Neelin, Jacob Bortnik, Raleigh Martin, Kate Ledger, Amato Evan, Shanna Shaked, and Ralph Kahn for helpful comments and discussions, and thank Philip Rosenberg for providing the data from Ref. [37]. We acknowledge support from National Science Foundation (NSF) grant 1552519 (J.F.K.), NASA grants NN14AP38G (D.A.R. and C.L.H.) and NNG14HH42I (R.L.M.), and from the U.S. Department of Energy as part of the Regional & Global Climate Modeling program (C.Z.).

**Author contributions**
J.F.K. conceived the project, designed the study, performed the analysis, and wrote the paper. D.A.R., C.Z., C.L.H., R.L.M., D.S.W., S.A., and K.H. contributed global model simulations. Q.Z. assisted with designing the statistical model to constrain dust properties from different data sets. All authors discussed the results and commented on the manuscript.

**Competing financial interests**
The authors declare no competing financial interests.

## Materials and Methods
**Analytical framework for constraining the size-resolved atmospheric dust loading.** Past constraints on the global dust loading and the resulting dust radiative effects have been obtained mostly from ensembles of global model simulations[4-6]. To simulate dust loading, these models must represent non-linear small-scale processes, such as dust emission and deposition[38], which are not resolved within large-scale climate models. These small-scale processes are thus heavily parameterized[39-41], introducing uncertainty in the simulated dust loading. In addition, model results can contain biases that arise from inconsistencies of assumed dust properties with respect to experimental and observational constraints[8, 9].

To overcome these limitations of global model ensembles, we have developed an analytical framework that constrains the global dust loading and its direct radiative effect using observational and experimental constraints, where available, to replace modeling results. Further, our framework directly links the global dust loading to a strong observational constraint on the magnitude of the global dust cycle: satellite measurements of the aerosol optical depth, which can be partitioned between that arising from dust and from other aerosols[4, 16, 42]. The dust aerosol optical depth (DAOD), which quantifies the extinction of solar radiation by dust, is constrained globally by years of retrievals from multiple satellites that have been calibrated against accurate ground-based measurements[43]. The global atmospheric loading of $PM_{10}$ dust ($L_{atm}$) can thus be expressed as,

$$L_{atm} = A_{Earth} \frac{\tau_d}{\varepsilon_\tau}, \tag{1}$$

where $A_{Earth}$ is the area of the Earth, $\tau_d$ is the globally-averaged DAOD at 550 nm wavelength, and $\varepsilon_\tau$ (m$^2$/kg) is the mass extinction efficiency. We use the results of Ridley et al.[16], who combined satellite measurements, ground-based measurements, and global transport model simulations to constrain the global DAOD to $\tau_d = 0.030$ (0.020 – 0.040) (Fig. 1a).

The globally-averaged mass extinction efficiency $\varepsilon_\tau$ equals the summed projected surface area of a unit mass of dust loading, multiplied by the globally-averaged efficiency with which a unit projected dust surface area extinguishes radiation. Because these factors depend on the dust geometric diameter $D$ (i.e., the diameter of a sphere with the same volume as the irregular dust particle), the contribution of each dust particle size to $\varepsilon_\tau$ must be weighted by the globally-averaged volume size distribution of atmospheric dust, $\frac{dV_{atm}}{dD}$, which is normalized (i.e., integrating over $D$ yields unity). That is,

$$\varepsilon_\tau = \int_0^{D_{max}} \frac{dV_{atm}}{dD} \frac{A(D)}{M(D)} Q_{ext}(D)\, dD, \tag{2}$$

where $A(D)/M(D) = 3/2\rho_d D$ is a spherical particle's projected surface area per unit mass, $\rho_d = (2.5 \pm 0.2) \cdot 10^3$ kg/m$^3$ is the density of dust aerosols (see Supplement); and $D_{max} = 20$ µm is the diameter above which the contribution to the global DAOD can be neglected, as justified by our results (Fig. S1). We further define the globally-averaged extinction efficiency $Q_{ext}(D)$ as the extinction cross-section normalized by $\pi D^2/4$, the projected area of a sphere with diameter $D$. Since an irregular dust particle has more surface area than a spherical particle with the same volume, it will generally have a larger extinction efficiency[18].

The globally-averaged size distribution of atmospheric dust, $\frac{dV_{atm}}{dD}$, is determined by three factors: (i) the normalized volume size distribution at emission ($\frac{dV_{emit}}{dD}$), (ii) the globally-averaged size-resolved dust lifetime ($T(D)$), and (iii) any changes in the size of dust particles during transport due to chemical processing and aggregation with other aerosols, which is likely insignificant for African dust[44, 45] but might play a role for Asian dust[46]. Such changes in dust size during transport are neglected in many models due to a lack of mechanistic understanding[3, 11, 35, 39, 47]. By similarly neglecting this process, we obtain

$$\frac{dV_{atm}}{dD} = \frac{dV_{emit}}{dD} \frac{T(D)}{\bar{T}}, \tag{3}$$

where the mass-weighted average dust lifetime $\bar{T}$ is given by

$$\bar{T} = \int_0^{D_{max}} \frac{dV_{emit}}{dD} T(D) dD. \tag{4}$$

where we have used the fact that both the atmospheric and emitted volume size distributions are normalized; note that $\bar{T}$ is also equal to $L_{atm}/F_{emit}$, where $F_{emit}$ is the global dust emission rate. The above equations yield $\varepsilon_\tau = 0.67$ (0.55–0.84) m²/g for PM$_{20}$ dust, which is consistent with results from the AeroCom global model ensemble[6]. We use $\varepsilon_\tau$ to obtain the size-resolved global dust emission rate and loading (Fig. 2 and 3).

We use these constraints on the size-resolved dust loading to similarly constrain the dust direct radiative effect, $\zeta$. Since $\zeta$ is generated by extinction of radiation by dust, it can be expressed as the product of the dust optical depth and the radiative effect produced per unit of optical depth[15],

$$\zeta = \int_0^{D_{max}} \frac{d\tau_d}{dD} \Omega(D) dD = \frac{L_{atm}}{A_{Earth}} \int_0^{D_{max}} \frac{dV_{atm}}{dD} \frac{A(D)}{M(D)} Q_{ext}(D) \Omega(D) dD, \tag{5}$$

where we used Eqs. (1) and (2) to write $\frac{d\tau_d}{dD}$ in terms of the dust size distribution and extinction efficiency. The radiative effect efficiency $\Omega(D)$ is the all-sky DRE that dust of diameter $D$ produces per unit DAOD. It depends on numerous properties of the Earth system, including the spatial and temporal variability of dust, the surface albedo, the vertical temperature profile, the distribution of radiatively-active species such as clouds and greenhouse gases, and the asymmetry parameter and single-scattering albedo of dust. The value of $\Omega(D)$ is thus not readily amenable to an analytical treatment, such that we use results from four global model simulations to estimate $\Omega(D)$ (see Supplementary Figure S3 and Supplementary Text).

We used a procedure similar to Eq. (5) to calculate the dust DRE that results from the atmospheric dust size distributions in AeroCom models (colored lines in Fig. 2b), for which we obtained the global extinction of atmospheric radiation as a function of dust size by combining the AeroCom dust size distributions (Fig. 2b) with the Mie theory extinction efficiency (brown line in Fig. 1b) assumed in AeroCom models[5,15,19,20,26] (see Supplement for additional details).

Our analytical framework has important limitations. First, our results rely on the constraint on global DAOD from Ref. [16] (Fig. 1a), which is consistent with both AeroCom model simulations[6] and with the MERRA Aerosol Reanalysis product[16]. Nonetheless, the analysis in Ref. [16] is subject to various possible biases, including due to the cloud-screening algorithm[48], due to the separation of dust optical depth from that of all other aerosols, due to the remotely-sensed optical depth retrieval algorithm for aspherical particles[49], and due to systematic differences between remotely-sensed clear-sky aerosol optical depth and all-sky optical depth. The uncertainty due to many, but not all, of these biases were quantified in Ref. [16], and have been propagated into the results presented here. Second, as is the case in many global models[3, 11], our analytical approach to constraining the size-resolved dust loading cannot explicitly account for changes in optical properties and size distribution during transport due to chemical processing, internal mixing with other aerosols, and absorption of water vapor[47, 50]. However, our methodology does implicitly account for some of the effects of internal mixing because the globally-averaged dust extinction properties are based on both fresh and aged dust from a range of source regions (see Supplement). Third, our constraint on the dust extinction efficiency uses numerical modeling results in which dust is represented as an ensemble of tri-axial ellipsoids[17]. This shape is an imperfect representation of the highly heterogeneous and mineralogy-dependent shape and roughness of real dust, and thus might produce systematic errors[18]. Further, the shortest axis (height) of these ellipsoids is poorly constrained due to a scarcity of measurements[23], which also prevent the propagation of uncertainty in the particle height distribution (see Supplement). We thus likely underestimate the uncertainty on the dust extinction efficiency. Fourth, our analytical framework uses globally-averaged properties of dust to calculate the global size-resolved dust loading and resulting dust radiative effects. The neglect of regional heterogeneity in dust properties could introduce errors by not accounting for covariance between dust properties. An example of this would be if the index of refraction or shape of dust depended substantially on particle size. However, experimental results suggest such covariances are small[51, 52]. Fifth, our constraints on the global dust DRE at TOA (Fig. 4) rely on an ensemble of four global model simulations of the size-resolved dust DRE (Fig. S3). These models assume specific optical properties that, although broadly consistent with remote sensing and in situ measurements (see Supplement), are not subject to the detailed experimental constraints that we have used for constraining the emitted dust size distribution and extinction efficiency. Sixth, our constraints likely underestimate the warming effect of LW scattering interactions, which are not accounted for in most global models. We therefore follow the treatment of Miller et al.[3], which is the only global modeling study that we are aware of that has accounted for the contribution of LW scattering to the dust DRE. Specifically, we assume that the DRE from LW scattering equals 30% of that produced by LW absorption. Since the DRE from LW scattering is likely of similar magnitude to that arising from LW absorption interactions[53], our constraint on the LW DRE should be seen as conservative.

A final limitation of our approach is that it is currently impossible to observationally constrain the globally-averaged dust lifetime. Consequently, we rely on an ensemble of model results (Fig. 1d), which could contain systematic biases. Since there are few observational constraints to test deposition schemes in models[40, 42], the uncertainty of dust lifetime might be incompletely represented. Further, some models underestimate the prevalence of coarse dust far from source regions[1, 16, 39], which could be partially explained by the fine-size bias in models (Fig. 2). However, this underestimation of coarse dust can also be due to processes missing from models, such as aggregation during transport, numerical errors in the size distribution treatment, the neglected effect of asphericity on dust settling, electrostatic charging, or errors in the (dry) deposition parameterization[32, 54, 55]. Such systematic biases towards underrepresentation of long-range coarse dust transport could have caused our results to underestimate the global dust emission loading. However, this would strengthen our conclusions that dust loading is slightly underestimated, that atmospheric dust is coarser than represented in current models, and that the dust DRE is more positive than accounted for in current models.

**Constraining the globally-averaged size-resolved shortwave extinction efficiency.** The extinction efficiency of the global population of dust particles depends on (i) its average real refractive index, (ii) its average imaginary refractive index, and (iii) the distribution of dust particle shapes. Based on extensive measurements, we take the globally-averaged real index of refraction at 550 nm as $n = 1.53 \pm 0.03$ (see Supplement). The uncertainty in the imaginary index of refraction $k$ is substantially larger, partially due to regional variations in shortwave-absorbing minerals like hematite[12, 13, 56]. However, since absorption accounts for only a small fraction of the total extinction, its influence on our constraint on the extinction efficiency (Fig. 1b) is limited. We take $k$ as a lognormal distribution with $\log(-k) = -2.5 \pm 0.3$ (see Supplement). Finally, measurements and theory indicate that the distribution of dust shapes in the atmosphere can be represented as tri-axial ellipsoids[17] with a height-to-major axis ratio of $\varepsilon_h = \sim 0.333$[23, 57], and a deviation of the aspect ratio from 1 (spherical) described by a lognormal distribution[51] with a median aspect ratio of $\bar{\varepsilon}_a = 1.7 \pm 0.2$ and a geometric standard deviation of $\sigma_{\varepsilon_a} = 0.6 \pm 0.2$. We converted these parameters to $Q_{ext}(D)$ using a dust single-scattering database[17]. Specifically, we assumed that each of these parameters is independent, and obtained a large number ($10^4$) of parameter sets ($m$, $n$, $\bar{\varepsilon}_a$, and $\sigma_{\varepsilon_a}$) by randomly choosing values from the probability distribution of each parameter. We used the resulting sets of values for $Q_{ext}(D)$, obtained from the single-scattering database[17], to obtain the median and CI (dashed line and shading in Fig. 1b). We calculated the extinction efficiency of spherical dust with identical index of refraction using Mie theory[58] (brown line in Fig. 1b).

**Constraining the globally-averaged dust size distribution at emission.** We interpreted each of the five emitted dust size distribution data sets[37, 59-64] as a measure of the globally-averaged size distribution of emitted dust. We did so because (i) differences between measurements from different soils within a given study are very small[37, 59-61], implying that differences in the emitted dust size distribution between different soils are relatively small[8], and (ii) the wind speed at emission has no statistically significant influence on the size distribution of emitted $PM_{10}$ dust[65]. These observations from dust flux measurements are supported by the invariance of in situ dust size distributions to source region[66] and wind speed[67]. We fit each of the five data sets[37, 59-64] with an analytical form derived from brittle fragmentation theory[8]. We then combined these five analytical functions representing each data set in a statistical model to obtain the maximum likelihood estimate (MLE) of the globally-averaged emitted dust size distribution (dashed line in Fig. 1c). We obtained the uncertainty (shaded area in Fig. 1c) using a modified bootstrap procedure. See Supplement for additional details.

**Constraining the globally-averaged dust lifetime.** We constrained the globally-averaged and size-resolved dust lifetime using an ensemble of global model results from previous studies[56, 68-71], supplemented with simulations from the global transport models WRF-Chem, GEOS-Chem, and HadGEM (see Supplement). We fit an exponential function to each of the nine simulation results, which we combined in a statistical model to obtain the MLE of the globally-averaged size-resolved dust lifetime. We obtained the uncertainty (shaded area in Fig. 1d) using a modified bootstrapping procedure. See Supplement for additional details.

**Analysis of AeroCom model simulations.** We used results from the Aerosol Comparison between Observations and Models (AeroCom) project[5, 6] as representative of the current generation of global models. We included the probability distributions of simulation results from these models in Figs. 1a and 3, which were obtained using kernel density estimation with a Gaussian kernel with standard smoothing parameter following equation (3.31) in Ref.[72]. Results from the 'median' AeroCom model were obtained by Ref.[6] by taking the median of each dust cycle variable for each grid box and month. AeroCom results in Fig. 3 from models that simulated a dust size range larger than $PM_{10}$ were corrected based on our constraints on the dust size distribution at emission (Fig. 1c) and in the atmosphere (Fig. 2a), respectively. Results from the subset of seven AeroCom models that reported the simulated dust size distributions (see Supplement) are included in Fig. 2. Some of these AeroCom models simulated a dust diameter range smaller than 20 μm, for which we similarly used our constraints to correct the normalized size distributions of atmospheric (Fig. 2a) and emitted (Fig. S2) dust to the $PM_{20}$ range.

**Code availability.** The codes used to conduct the analysis presented in this paper and in the production of the figures are available at https://github.com/jfkok/KokDustDRE2017

**Data availability.** Data used in Figures 1-3 are included in the code. The global climate and chemical transport model simulation data that were used to constrain the dust DRE in Figure 4 are available through Zenodo (*doi to be added when data is complete*).

**References only in Methods**